\documentclass[hidelinks]{article}

\usepackage{arxiv}

\usepackage[utf8]{inputenc} 
\usepackage[T1]{fontenc}    
\usepackage{hyperref}       
\usepackage{url}            
\usepackage{booktabs}       
\usepackage{amsfonts}       
\usepackage{nicefrac}       
\usepackage{microtype}      
\usepackage{graphicx}
\usepackage[numbers,sort&compress]{natbib}
\usepackage{doi}
\usepackage{enumitem}
\usepackage{caption}

\newcommand\blfootnote[1]{%
  \begingroup
  \renewcommand\thefootnote{}\footnote{#1}%
  \addtocounter{footnote}{-1}%
  \endgroup
}

\title{Asynchronous Sensor System\protect\\for Collecting Detailed Data\protect\\on the Environment and Resource Consumption\protect\\in Smart City
}

\author{ \href{https://orcid.org/0000-0002-3309-7326}{Sergey Surnov\hspace{1mm}\includegraphics[scale=0.06]{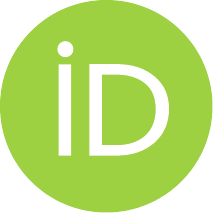}} \\
	\And
	\href{https://orcid.org/0000-0001-6927-0031}{Igor Bychkovskiy\hspace{1mm}\includegraphics[scale=0.06]{orcid.pdf}} \\
	\And
	\href{https://orcid.org/0000-0003-3408-3855}{Grigory Surnov\hspace{1mm}\includegraphics[scale=0.06]{orcid.pdf}} \\
	\And
        \href{https://orcid.org/0009-0007-1636-7586}{Nikolay Surnov\hspace{1mm}\includegraphics[scale=0.06]{orcid.pdf}}
}

\date{}

\renewcommand{\headeright}{}
\renewcommand{\shorttitle}{Asynchronous Sensor System for Collecting Detailed Data on the Environment and Resource Consumption}

\hypersetup{
pdftitle={Asynchronous Sensor System for Collecting Detailed Data on the Environment and Resource Consumption in Smart City},
pdfauthor={Sergey Surnov},
pdfkeywords={asynchronous sensor system, detailed data, smart meter, Smart City, resource consumption, meteorological parameters, air pollution, energy transition},
}

\begin{document}
\maketitle

\blfootnote{Sergey Surnov Ph.D., Igor Bychkovskiy, Grigory Surnov and Nikolay Surnov are with Divteh Ltd, Moscow, Russia.}

\blfootnote{\textit{Corresponding Author:} Sergey Surnov Ph.D., \href{mailto:surnov53@gmail.com}{\texttt{surnov53@gmail.com}}}

\begin{abstract}
This article expands on the ideas presented in \cite{s1}.

The article demonstrates that within a unified monitoring system, cities can collect not only detailed resource consumption data but also information on the environmental conditions under a common set of rules.

A method for constructing asynchronous sensor monitoring systems for controlled parameters in Smart City is proposed. The controlled parameters include: resource consumption in apartment buildings (electricity, cold and hot water, heat, gas); indoor and outdoor air pollution indicators (carbon monoxide, nitrogen oxides, hydrocarbons, dust, heavy metals, radiation levels, etc.); meteorological parameters (air temperature and humidity, atmospheric pressure, wind speed and direction). 

In an asynchronous sensor monitoring system, an event occurs when the value of a controlled parameter changes by a specified amount. This enables adjusting the granularity of the collected data. More detailed data contains more information and is therefore more valuable. 

Transitioning from traditional synchronous systems, where the value of a controlled parameter is recorded at set time intervals, to asynchronous systems allows for the abandonment of complex “smart meters” and the use of extremely simple and inexpensive sensors. Standardizing the data transmission protocol for all types of controlled parameters and reducing the cost of the most widespread equipment in the system — sensors — leads to lower expenses for creating and operating a monitoring system. 

Lower costs and increased value of collected data enable the potential opening of a new market — a market for data on resource consumption and environmental parameters in Smart City, attracting private businesses to this area and accelerating the energy transition towards a global economy with zero greenhouse gas emissions.

\end{abstract}


\section{Introduction}
In preparation for the global UN Climate Change Conference held in Glasgow from October 31 to November 12, 2021, the Intergovernmental Panel on Climate Change (IPCC), the World Meteorological Organization (WMO), the International Energy Agency (IEA) and the World Health Organization (WHO) prepared relevant estimates, projections and proposals related to global warming. Let us highlight the main ones, directly related to the topic of this article.

\begin{samepage}
Conclusions of the IPCC \cite{cc21}:
\begin{itemize}[nosep,topsep=0pt]
    \item the main cause of global warming is human activity;
    \item the best-case scenario is that carbon dioxide emissions must begin to be reduced immediately in order to reach zero after 2050;
    \item in the best-case scenario, the temperature increase will stop after 2050.
\end{itemize}
\end{samepage}

\begin{samepage}
Proposals of the IEA \cite{nz50}:
\begin{itemize}[nosep,topsep=0pt]
    \item it is necessary to increase energy efficiency by all available means: to reduce energy consumption by home heating and cooling systems, and to reduce energy losses in homes by modernizing them;
    \item people will have to change their behavior and learn to ensure resource conservation.
\end{itemize}
\end{samepage}

\begin{samepage}
Three large-scale initiatives of the WMO:
\begin{itemize}[nosep,topsep=0pt]
    \item the implementation of the \href{https://public.wmo.int/ru/%D1%80%D0%B5%D0%B7%D0%BE%D0%BB%D1%8E%D1%86%D0%B8%D1%8F-%D0%B2%D0%BC%D0%BE-%D0%BE-%D0%B5%D0%B4%D0%B8%D0%BD%D0%BE%D0%B9-%D0%BF%D0%BE%D0%BB%D0%B8%D1%82%D0%B8%D0%BA%D0%B5-%D0%B2-%D0%BE%D0%B1%D0%BB%D0%B0%D1%81%D1%82%D0%B8-%D0%B4%D0%B0%D0%BD%D0%BD%D1%8B%D1%85-%D0%B2%D0%BD%D0%B5%D0%BE%D1%87%D0%B5%D1%80%D0%B5%D0%B4%D0%BD%D0%B0%D1%8F-%D1%81%D0%B5%D1%81%D1%81%D0%B8%D1%8F-%D0%B2%D1%81%D0%B5%D0%BC%D0%B8%D1%80%D0%BD%D0%BE%D0%B3%D0%BE-%D0%BC%D0%B5%D1%82%D0%B5%D0%BE%D1%80%D0%BE%D0%BB%D0%BE%D0%B3%D0%B8%D1%87%D0%B5%D1%81%D0%BA%D0%BE%D0%B3%D0%BE}{WMO Uniform Data Policy} for the data \cite{wmo1} of observations;
    \item the creation of the \href{https://community.wmo.int/gbon}{Global Basic Observing Network} \cite{gbon1}, which is considered by the WMO as a new overarching structure for all observing systems;
    \item the creation of the \href{https://public.wmo.int/en/our-mandate/how-we-do-it/development-partnerships/Innovating-finance}{Systematic Observations Finance Facility} \cite{un1}.
\end{itemize}
\end{samepage}

In September 2021, the WHO published new Air Quality Guidelines \cite{who1}, which assess the impact of pollutants on human health.

The main conclusion is that atmospheric air pollution ranks fourth (out of 28) among the world’s risk factors for death. Harm is caused at lower concentrations of pollutants than previously thought. The WHO has adjusted nearly all limit values for major pollutants, for instance, reduced the average annual safe concentration of fine particles in the air by half, and that of coarse particles, by 25\%.

\begin{samepage}
As observed, the focus is on three interconnected factors that significantly influence greenhouse gas emissions and living conditions:
\begin{itemize}[nosep,topsep=0pt]
    \item resource consumption (electricity, cold and hot water, thermal energy, and gas);
    \item weather conditions;
    \item atmospheric air pollution.
\end{itemize}
\end{samepage}
\begin{samepage}
Carrying out these initiatives entails:
\begin{itemize}[nosep,topsep=0pt]
    \item a significant expansion of data collection on environmental conditions and resource consumption worldwide;
    \item broad application of scientific and technological achievements in data collection, storage, and processing;
    \item enhancing the capabilities and scale of private sector activities in meteorology and resource conservation.
\end{itemize}
\end{samepage}

\subsection{Resource conservation potential}
Research conducted within the UN Human Settlements Program \cite{gw19} revealed that over 80\% of global energy resources are consumed in urban areas. Consequently, cities are responsible for 70\% of greenhouse gas emissions.

In cities, the population is one of the main end users of resources (cold and hot water, electricity, thermal energy, and gas). For instance, data from \cite{antonov20} shows that more than 55\% of the final demand for resources is generated by the population of a large European metropolis. The population consumes over 40\% of thermal energy and more than 20\% of electricity. This paper also highlights the significant influence of weather on the consumption of thermal energy, electricity, and gas, as well as their proportion in the city’s fuel and energy balance. The production of these types of resources directly impacts the pollution of atmospheric air.

What is the potential for reducing resource consumption and, consequently, decreasing greenhouse gas emissions in cities?

\begin{figure}
    \centering
    \captionsetup{justification=centering,margin=2cm}
    \includegraphics[width=.9\linewidth]{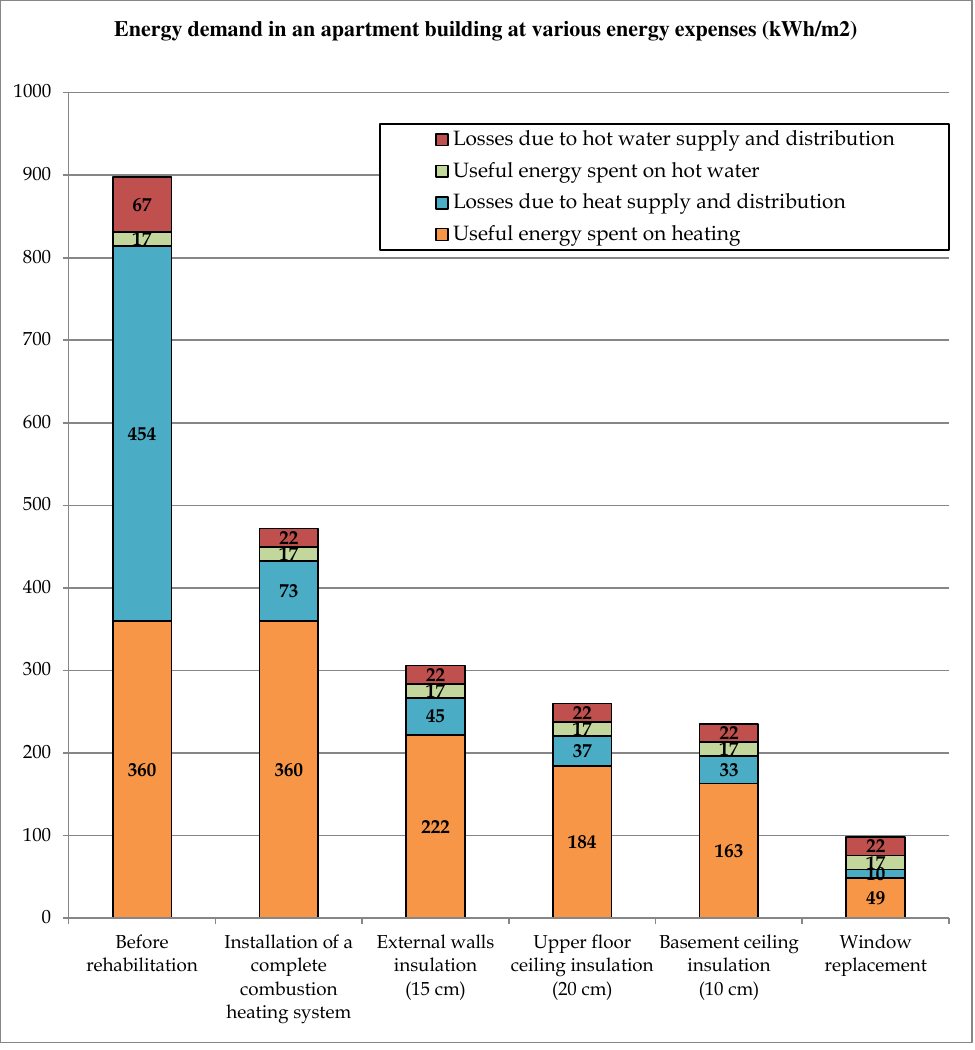}
    \caption{Energy saving potential in a city (example)\protect\\
(according to TÜV Rheinland Group (Cologne) and the Institute for Housing and Ecology (Darmstadt)}
    \label{fig:potential}
\end{figure}

The Russian-German study “Korolev as a Model City for Heat Conservation” \cite{linke05}, conducted in 2005-2006 at the request of the German Ministry of Economics, demonstrated a significant potential for heat savings in Russian cities (Figure \ref{fig:potential}). 

The main conclusion is that the consumption of thermal energy by the population (and consequently, the use of gas for heating) can be reduced by 50-90\% (depending on the expenses for building modernization). Assuming a reduction of at least 50\%, atmospheric emissions would accordingly decrease by 58\%. This not only contributes to the reduction of the greenhouse effect but also to the decrease in atmospheric air pollution. The study also emphasized the necessity of collecting detailed data on the consumption of each type of resource by each consumer as a foundation for evaluating the effectiveness of resource-saving measures. 

While resource consumption traditions vary among countries, this example only reinforces the IEA’s initiative to immediately reduce resource consumption and losses.

\subsection{``Smart meters'' and the detailed data market}
With the emergence of “smart meters” of resources \cite{smwiki}, there is much anticipation for their rapid proliferation and the development of a new market — the market for detailed resource consumption data. Indeed, sufficiently detailed data provide extensive information about the structure of resource consumption, consumers of the resources, their habits, the household equipment they use, the state of infrastructure, and so on. This information can be utilized for resource conservation purposes.

As a result, a separate research direction and numerous publications have emerged on the topic of NILM (Nonintrusive Load Monitoring) \cite{hart92}, a method for analyzing aggregated electrical load data obtained by measuring current and voltage at a single point, followed by dividing the total load among individual devices. This approach is effective for analyzing the structure of electricity consumption at enterprises. When applied on a large scale to analyze the urban population’s power consumption, issues arise, mainly due to the cost of equipment needed for collecting detailed data. 

Several large-scale projects involving “smart meters” were initiated, but their success was later doubted.

In 2009, Google introduced the Google PowerMeter platform \cite{kopytoff09} aimed at collecting detailed data on electricity usage and promoting energy conservation among consumers. It was expected that “smart meters” would be installed worldwide within a few years, reducing electricity consumption by up to 15\%. However, in September 2011, the platform’s operations were terminated due to “scaling issues,” meaning the “smart meter” adoption forecast did not materialize \cite{pm11}.

From 2011 to 2018, the UK government spent 11 billion pounds sterling on a “smart meter” installation program. Interestingly, the Foundation for Information Policy Research initially criticized \cite{anderson10} the program, arguing that it would unlikely reduce electricity consumption, was costly, and did not foster competition. They pointed out that suppliers and consumers have opposing interests: suppliers aim to sell more electricity, while consumers want to use less. A suggestion was made that resource consumption tracking should be managed by an independent operator or competing operators. These observations remain relevant today.

J’son \& Partners Consulting’s studies “The status and prospects for the use of LPWAN radio technologies in various segments of the Internet of Things (IoT) market” (August 2017) \cite{lpwan17}, and “The Russian market of machine-to-machine communications and the Internet of Things at the end of 2019, forecast until 2025” (July 2020) \cite{rumarket20} highlight that the main obstacle to adoption of “smart meters” is their high cost and the lack of successful business models. Connecting “smart meters” becomes economically unfeasible (the cost of obtaining data exceeds the value of this data) and can only be implemented through administrative, rather than market, methods. There is no market for data on resource consumption in Smart City.

There is a reason for this situation.

\subsection{Resource suppliers and consumers}
It should be noted that resource suppliers and consumers attribute different meanings to the term “resource conservation.”

The suppliers refer to reducing resource losses during its delivery to the consumers via the corresponding infrastructure. This reduces the costs of the suppliers. The development of smart grids \cite{sgwiki} in the electric power industry is a vivid example of the suppliers’ efforts to enhance infrastructure manageability, decrease operational expenses, and minimize resource transportation losses. Smart grids entail collecting detailed data about the infrastructure, resource transportation processes, and control actions on objects within the infrastructure to optimize specific parameters. The suppliers are motivated to achieve this goal and have the capability to progress towards its realization. They are aware that the expenses for smart grids will be compensated by reducing resource losses and infrastructure operation costs. However, the suppliers do not intend to decrease the consumers' resource consumption. On the contrary, they are interested in having the consumers purchase as many resources from them as possible.

The consumers, by using the term “resource conservation,” mean reducing the amount of resources they consume and, consequently, cutting their expenses. While reducing resource transportation losses and infrastructure operation costs is an important objective, consumer behavior does not influence its achievement. Equally important is to encourage people to conserve resources and demonstrate how to do so. Solving this issue requires obtaining detailed data on each resource’s consumption by every consumer, which directly contradicts the interests of the suppliers.

Assigning various additional functions to resource meters (such as storing and transmitting accumulated data on quantity and quality of consumed resources, synchronizing time within the system, controlling unauthorized access, and so on) has significantly complicated these meters, leading to them being called “smart.” As a result, the production cost of the meters and their operational expenses within the system have increased. No one has set the goal of obtaining sufficiently detailed data from them. These meters are designed for the suppliers, not the consumers.

It should also be noted that currently, each resource’s consumption is tracked by a separate data collection system, independent of others, and environmental parameter monitoring in these systems is not considered at all.

\subsection{Environmental parameter monitoring}
Regarding systems that collect meteorological data, atmospheric pollution data, and radiation situation data, these are synchronous systems where sensors are polled at specified intervals. The WMO has concluded that the number of data collection points on Earth is evidently insufficient. Increasing their quantity, along with enhancing the granularity of collected data, will contribute to the improvement of climate models and the accuracy of environmental forecasts.

\section{Monitoring system as an instrument for energy transition}
Considering cities’ significant contribution to the greenhouse effect and the initiatives of the IEA, WMO, and WHO regarding energy transition, the main objectives of creating a monitoring system in Smart City can be listed as follows:
\begin{enumerate}[nosep,topsep=0pt]
    \item the system should focus on collecting detailed data about the consumption structure of all resources by end users (primarily the population) and their losses;
    \item given the connection between resource consumption and the consumer’s environment, detailed data should also be collected on meteorological parameters and indoor and outdoor air pollution levels;
    \item expanding the capabilities and scale of private sector involvement in environmental and resource conservation (establishing a data market), specifically:
\begin{itemize}[nosep,topsep=0pt]
    \item enhancing the value of collected data by increasing their granularity;
    \item reducing the expenses for development and operation of a monitoring system by implementing innovative technical solutions;
\end{itemize}
    \item maintaining the monitoring system’s properties during scaling, including providing an acceptable level of security.
\end{enumerate}

\section{Transitioning to an asynchronous sensor-based monitoring system}
\subsection{Synchronous systems}
Conventional systems for collecting data on resource consumption, weather, and air pollution are synchronous, meaning that readings from resource meters and controlled parameter sensors are taken at specified time intervals (polling intervals).

A typical example of a synchronous system is an AMI system for collecting resource consumption data using “smart meters” (the AMI is the Advanced Metering Infrastructure \cite{sggloss}). In such a system, an event occurs when a predetermined time interval (polling interval) has passed. The response to the event includes recording meter readings and their corresponding times, storing accumulated readings, transmitting them to the Monitoring Center, and so on.

It is worth noting that these actions also take place when there is no resource consumption. The granularity of collected data can be adjusted by modifying the polling interval. Meter design is complicated by the need for time synchronization within the system, monitoring the meter’s condition, and implementing measures to prevent unauthorized access.

As a result, the cost of “smart meters” and operational expenses increase.

The most promising and actively developed interfaces in an AMI system between meters and the Monitoring Center are NB IoT (Narrow Band Internet of Things) and LoRaWAN (Long Range Wide Area Networks).

In an AMI system, replacing or modifying the data interface built into the meter will require it to be re-certified as a measuring instrument. An AMI system is sensitive to scaling. This must be taken into account when the number of meters in the system grows.

As shown in [1], the transition from synchronous systems to asynchronous systems (the Smart Monitoring technology) avoids a significant number of problems that arise when using the AMI systems to obtain sufficiently detailed data on the consumption of resources by the population in Smart City.

\subsection{Asynchronous systems}
The first goal of moving to asynchronous systems is to obtain data on changes in controlled parameters at a level of the granularity that will allow us to extract the information of interest and thereby increase the value of the data being collected.

The systems that implement the Smart Monitoring technology will be referred to as the ASMI (Asynchronous Sensory Monitoring Infrastructure) systems.

\begin{figure}
    \centering
    \captionsetup{justification=centering,margin=2cm}
    \includegraphics[width=.9\linewidth]{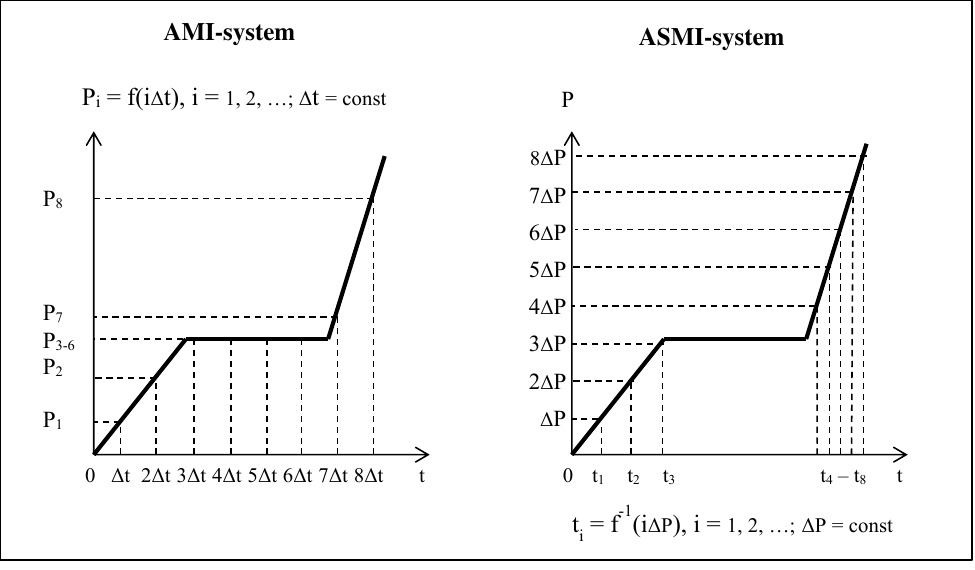}
    \caption{Event response in AMI-system and ASMI-system}
    \label{fig:amiasmidiff}
\end{figure}

The response to an event in an AMI system and an ASMI system is illustrated in Figure \ref{fig:amiasmidiff} which shows the dependence of the amount of resource \emph{P} consumed on time \emph{t}.

In an AMI system, an event occurs when a specified time interval $\Delta t$ has elapsed. The response to an event consists in recording the value of the controlled parameter \emph{P}. The degree of the granularity of the collected data is regulated by the change of $\Delta t$.

In an ASMI system an event occurs when the value of the controlled parameter changes by the set value $\Delta P$. The response to an event consists in recording the time of the event and its sequence number. The degree of the granularity of the collected data is controlled by the change of $\Delta P$. In this case, if the value of the controlled parameter does not change (for example, the resource is not consumed), then the event and, accordingly, the response to it do not occur.

Both systems may additionally record some parameters related, for example, to the diagnostics of the system state.

The second objective of transitioning to asynchronous systems is to decrease the expenses associated with constructing and maintaining the ASMI systems, while ensuring their scalability and an acceptable level of security.

In order to achieve this objective, routers have been incorporated into the structure of the ASMI systems as an intermediary link between the controlled parameter sensors and the Monitoring Center. The structure of an ASMI system is depicted in Figure 3.

The components of an ASMI system are:
\begin{enumerate}[nosep,topsep=0pt]
    \item Controlled parameter sensors. They are installed in apartment buildings, both inside and outside the apartments, depending on the parameters being controlled.
    \item Routers. They are positioned within a building but outside the apartments, ensuring that each sensor is within radio range of at least one router. They are mains operated.
    \item The Monitoring Center where all messages transmitted by routers are collected and processed.
\end{enumerate}

\subsection{Controlled parameter sensors}
The sensors are the most abundant devices within an ASMI system. The overall cost of creating and operating the entire system is largely influenced by their complexity and corresponding price. Hence, during the development of the Smart Monitoring technology, a particular focus was placed on reducing “intelligence” of the sensors and minimizing the number of functions they perform. 

A sensor has two roles: it detects the occurrence of an event and, upon its occurrence, transmits a message containing the unique sensor number, the sequence number of the transmitted message, and additional data crucial for operation of the system. A sensor always initiates communication. All routers within radio range receive transmitted messages. Communication is one-way, from a sensor to a router, using the standard Pi protocol for all controlled parameters. A sensor keeps track of the number of messages it has transmitted (communication sessions). A sensor’s number uniquely defines the structure of its transmitted message, enabling the Monitoring Center to extract the information within the message.

For diagnostics, a sensor sends a status control message at specified time intervals.

In an ASMI system, a display on a sensor is not mandatory.

\subsection{Routers}
A router’s function is to receive a message from a sensor, add the message’s receipt time, and forward it to the Monitoring Center. Messages from sensors to routers are sent using the Pi protocol. A router can connect with the Monitoring Center through one of the currently popular interfaces (e.g., NB IoT, LoRaWAN, GPON) or another interface.

\begin{figure}
    \centering
    \captionsetup{justification=centering,margin=2cm}
    \includegraphics[width=.9\linewidth]{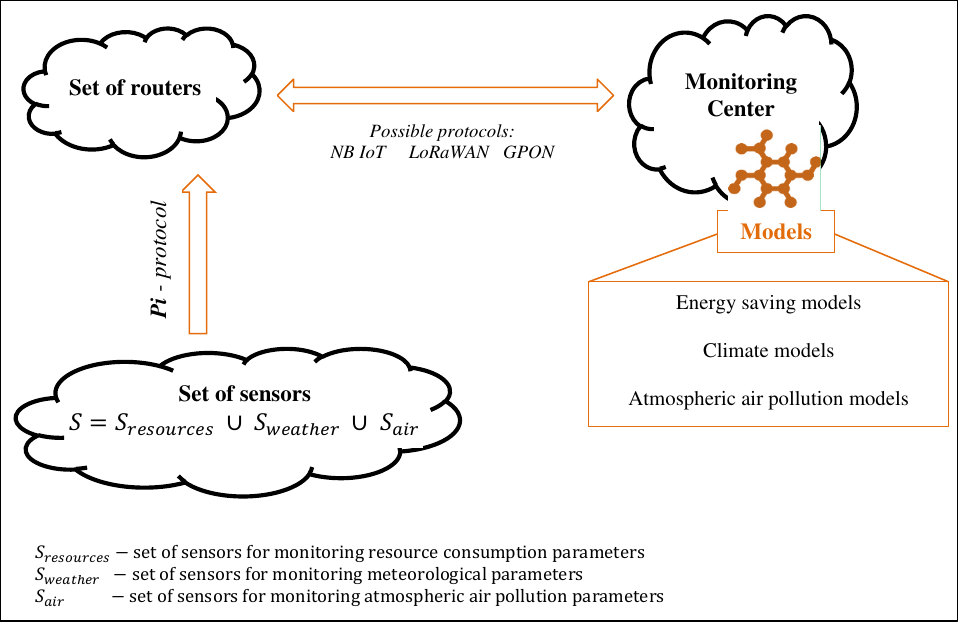}
    \caption{ASMI-system structure}
    \label{fig:asmistruct}
\end{figure}

An important feature of the ASMI systems for our article’s purpose is that routers do not differentiate between sensors. A router stores all incoming messages from all sensors within radio range and later forwards them to the Monitoring Center. The Monitoring Center, along with numerous routers in city apartment buildings, creates a transport network of the ASMI system, which is independent of the data type being transmitted.

\subsection{Monitoring Center}
The Monitoring Center stores data about the sensors included in the system, their installation locations, the parameters controlled by them, the set values $\Delta P$, etc. Data about routers included in the system and their installation locations are also stored here.

The Monitoring Center receives messages sent by routers. Here they are processed and stored for future use. The Monitoring Center synchronizes the time in routers. The resulting data can be used for resource conservation purposes, environmental assessment, settlements between suppliers and consumers of resources, etc.

Introduction of a new sensor into the system consists in placing it within radio range of at least one router and registering it in the Monitoring Center, and indicating at least its number, location, type of the controlled parameter, and the value $\Delta P$.

Thus, using the ASMI systems also for monitoring meteorological parameters (temperature and humidity, atmospheric pressure, wind speed and direction, radiation levels) and air pollution parameters inside and outside the premises in Smart City does not require any special modifications to the system. In this case, it is possible to control the level of the granularity in both collected data and geographical detail.

The invention related to the Smart Monitoring technology has been granted a patent in the US \cite{uspat20}.

\subsection{Advantages of the ASMI systems}
The advantages of the ASMI systems include the following:
\begin{enumerate}[nosep,topsep=0pt]
    \item All controlled parameters are monitored using a single technology;
    \item The level of the granularity in the data is set for each controlled parameter. Data is transmitted from sensors to routers only when the controlled parameters change by a given value;
    \item The design of the controlled parameter sensors is simplified to the maximum; in particular, the time synchronization function has been transferred from sensors to routers; 
    \item To increase reliability, routers receive messages from all sensors that are within radio range;
    \item To prevent unauthorized access to sensors, one-way communication from sensors to routers is used;
    \item The communication protocol between sensors and routers is unified for all controlled parameters. This prevents the need for re-certification of sensors as measuring instruments when upgrading interfaces used for data transmission from routers to the Monitoring Center;
    \item Open frequency ranges are used for communication between sensors and routers;
    \item An ASMI system easily scales to any size.
\end{enumerate}

\subsection{Models}
The flow of detailed, diverse, and loosely connected data from numerous sensors spread across a vast area implies the application of machine learning technologies.

Detailed data enables the construction of models and extraction of information contained within it in at least the following ways (the list is not exhaustive).
\begin{enumerate}[nosep,topsep=0pt]
    \item Resource consumption behavior of residents and the household equipment they use:
\begin{itemize}[nosep,topsep=0pt]
    \item identification of household appliances in each apartment and determination of their energy efficiency class;
    \item disaggregation of the total consumption of each resource by household appliances in an apartment;
    \item keeping a log of the operation of each appliance in an apartment (consumption, repairs, etc.);
    \item recommendations to consumers on the use of more energy-efficient household appliances;
    \item recommendations to residents on the change of their behavior in order to ensure resource conservation;
    \item determining the actual number of people living in an apartment or a building; analyzing the population density in an area, and tracking its changes over time;
    \item support of settlements between suppliers and consumers of resources; recommendations to consumers on the use of multi-tariff meters.
\end{itemize}
    \item Energy efficiency of housing and ways to enhance it:
\begin{itemize}[nosep,topsep=0pt]
    \item determination of the energy efficiency class of an apartment and a building as a whole; 
    \item preparation of recommendations for measures to improve the energy efficiency of residential buildings;
    \item evaluation of the effectiveness of the measures taken for resource conservation.
\end{itemize}
\begin{samepage}
    \item Environmental predictions:
\begin{itemize}[nosep,topsep=0pt]
    \item forecasting of climate change and preparation of recommendations for changes in the fuel and energy balance of an area;
    \item forecasting of the state of the environment (air pollution, radiation situation, meteorological conditions);
    \item detection and containment of natural disasters and emergencies, forecasting their spread throughout an area and assessment of possible damage.
\end{itemize}
\end{samepage}
\end{enumerate}

\section{Conclusion}
The authors hope that the ASMI systems will become a convenient tool for facilitating the energy transition by:
\begin{itemize}[nosep,topsep=0pt]
    \item jointly monitoring resource consumption, meteorological parameters, and indoor and outdoor air pollution parameters;
    \item adjusting the level of the granularity in the collected data, which allows for increasing their value;
    \item reducing the costs associated with creating and operating an ASMI system.
\end{itemize}

Collectively, this creates favorable conditions for establishing a new market — a data market, and consequently, for attracting private investment into resource conservation and environmental forecasting sectors.

\end{document}